\documentclass[aps,prb,twocolumn,showpacs,longbibliography]{revtex4-1}
\usepackage{amssymb}
\usepackage{amsmath}
\usepackage{amsfonts}
\usepackage{graphicx}
\usepackage{color}
\usepackage[utf8x]{inputenx}
\usepackage{hyperref}
\usepackage{bm}

\begin{document}
\title{Electrically controlled crossing of energy levels in quantum dots in two-dimensional topological insulators}

\author{Aleksei~A. Sukhanov}
\affiliation{V.A. Kotel’nikov Institute of Radio Engineering and Electronics, Russian Academy of Sciences,
Fryazino, Moscow District, 141190, Russia}

\begin{abstract}

We study the energy spectra of bound states in quantum dots (QDs) formed by an electrostatic potential in two-dimensional topological insulator (TI) and their transformation with changes in QD depth and radius. It is found that, unlike a trivial insulator, the energy difference between the levels of the ground state and first excited state can decrease with decreasing the radius and increasing the depth of the QD so that these levels intersect under some critical condition. The crossing of the levels results in unusual features of optical properties caused by intraceneter electron transitions. 
 In particular,  it leads to significant changes of light absorption due to electron transitions between such levels and to the transient electroluminescence induced by electrical tuning of  QD and TI parameters. In the case of  magnetic TIs, the polarization direction of the absorbed or emitted circularly polarized light is changed due to the level crossing.

\end{abstract}


\maketitle

\section{Introduction}
\label{Intro}

 In the study of two-dimensional (2D)  time-reversal invariant topological insulators (TI), the main attention is focused on the gapless electronic states that arise at the edges of 2D  TI or at boundaries of  the TI with conventional insulators ~\cite{QiZhangRMP2010,HasanKaneRMP2010,Shen1}. These states have linear Dirac spectrum and helical spin structure.  Due to spin–momentum locking they are robust against elastic scattering by nonmagnetic defects and form one-dimensional channels which have quantized conductance and carry the equilibrium spin current. These features of the edge states make them promising for the use in spintronics. In this regard, the possibility of the electrical control of charge and spin currents were extensively investigated for a variety of nanostructures based on 2D TIs, such as quantum point contacts ~\cite{Dolcetto,Romeo} and tunnel junctions with conventional semiconductors ~\cite{SS}.

 Another interesting aspect of the 2D TI study is due to  fundamental changes in energy spectrum and spin structure of defects and quantum dots which are caused by strong spin-orbit coupling and inverted band structure ~\cite{Shan1,Lu1,Sabl,Slager,QiZhang,Mesaros}.
 
In Ref. ~\onlinecite{Shan1}, it was shown that the bound states in 2D TI occur at uncharged vacancies (antidotes).  Basically, these states are edge states localized at the perimeter of an antidot.   The emergence of in-gap bound states induced by impurities with a Gaussian potential was  demonstrated in Ref. ~\onlinecite{Lu1} by numerical calculations for a certain set of TI parameters. 

 In the work ~\onlinecite{Sabl} it was shown that defects with the short-range potential in 2D TI bulk, unlike the trivial semiconductor, can host at once two bound states of different nature. The states of first kind are due to the localization of carriers for which the defect potential is attractive, while other states are akin to edge states and caused by the localization of carriers, for which the defect potential is repulsive.

 Interestingly, the localized states in 2D TI may occur at magnetic vortices ~\cite{QiZhang,Mesaros}. Thus in Ref. ~\onlinecite{Mesaros} the existence of a pair of zero-energy modes bound to a vortex carrying a $\pi$-flux was demonstrated.

Note , that in recent years the electron states localized at different kinds of defects in three-dimensional (3D) topological  insulators, in particular, impurity-induced states on the surface of 3D TI ~\cite{Biswas} and the states localized  on the linear dislocations  ~\cite{Imura}  and grain boundaries ~\cite{RJS} were also  intensively studied .

 From both practical and pure scientific points of view, quasi-zero-dimensional quantum dots (QDs) with controlled parameters are of great interest  ~\cite{Chakra}. 

Quantum dots in 2D TI have attracted attention primarily because of the appearance of the circular edge states with quantized spectrum on their borders. Usually, the isolated QDs with an infinite barrier at their borders are  studied. ~\cite{Isaev,Chang,Li}
Besides the ordinary size-quantized states with energy levels located in the conduction and valence bands, these topological QDs support also the "edge" bound states with levels laying in the  energy gap.
 
In the work ~\onlinecite{Herath} the spectra of intrasubband and intersubband optical transitions between localized states in QD created  in an ultra-thin film 3D TI are calculated. In  such film the energy gap exists in the spectrum of surface states exists. The local increase in the film thickness leads to the reduction of the energy gap and the occurrence of quantum wells at once for both electrons and holes.

Here we study the properties of QDs formed in 2D TI 
by the strong potential, comparable to or larger than the energy gap. 
We are interested in issues how the unusual properties of topological insulators will manifest themselves in electrostatically confined quantum dots, where the  trivial phase region is absent.
 We focus on the properties of the deep QDs of small radius. It is essential that such quantum dots can be considered as model defects with a strong short-range potential. Recall that the point defect in the bulk of 2D TI can host two bound states in contrast to the trivial case where only one state can exist.~\cite{Sabl} So our study aims to trace the evolution of the bound states with a decrease in the QD size and identify opportunities  for the two states  coexistence in small QD, similar to the case of point defect.

We show that in contrast to the case of trivial 2D insulator, in TI  quantum dot with a strong short-range potential, the energy gap between ground and excited states can decrease with decreasing radius and  increasing depth of the QD. 
Furthermore, in QDs of fairly small radius the relative position of the energy levels of these states can be inverted. 

This can lead to interesting effects including the strong variation in the intensity of the absorbed  light. Moreover, in case of magnetic TI the inversion of the levels  results in the direction change of the circular polarization of  light absorbed while passing through the 2D TI.

In Section II we describe the model and the calculation scheme. In Section III the calculation results are presented and the possible experimental manifestations of level crossing are discussed. At last in Section IV
we qualitatively explain the transformations of the bound state spectra      
and summarize the main results.

\section{The model and calculation scheme}\label{model}

Consider a 2D TI with QD formed by a gate located above it. The radius and the depth of the potential well in the QD are controlled by the gate potential.
To describe the bound states in the QD, we use the effective model Hamiltonian  proposed by Bernevig, Hughes and Zhang  (BHZ)  ~\cite{BHZ} for 2D TI:
\begin{equation}
 H=
\begin{pmatrix}
 h(\mathbf{k}) & 0\\
 0 & h^*(-\mathbf{k})
\end{pmatrix}\,,
\label{H}
\end{equation} 
where $\mathbf k$ is momentum operator and
\begin{equation}
h(\mathbf{k})=
\begin{pmatrix}
 M\!+\!(B\!+\!D)\mathbf{k}^2 & \mathcal{A}(\mathbf{k_x}+i\mathbf{k_y})\\
 \mathcal{A}(\mathbf{k_x}-i\mathbf{k_y}) & -M\!-\!(B\!-\!D)\mathbf{k}^2 
\end{pmatrix}\,.
\label{h}
\end{equation} 

The Hamiltonian  ~(\ref{H}) describes the electron states of  2D TI expressed in the basis of $\{|E_1\uparrow\rangle, |H_1\uparrow\rangle,|E_1\downarrow\rangle,|H_1\downarrow\rangle\}$, where $\{|E_1\uparrow\rangle,|E_1\downarrow\rangle,\}$ are the superposition of the electron states of s-type and the light-hole states of p-type with spin-up ($\uparrow$) and spin-down ($\downarrow$) and  $\{|H_1\uparrow\rangle,|H_1\downarrow\rangle\}$ are the heavy-hole states. 

In Eqs.~(\ref{h}) the model parameters $(\mathcal{A}, B, D)>0$ and $M<0$ in the topological phase while $M>0$ in trivial phase.

The BHZ model is commonly used to study the transport, optical and photoelectric properties of edge states ~\cite{Sonin,Aseev,Entin}, bound states of defects and impurities ~\cite{Shan1,Lu1,Sabl,Sablik}, localized states in the line dislocations  ~\cite{Imura}  and on grain boundaries ~\cite{RJS} and QD states ~\cite{Isaev,Chang,Herath}. In the original form this model did not take into account the spin-orbit interaction (SOI) but further it was generalized to include the SOI due to the  bulk inversion asymmetry (BIA)~\cite{Koenig}  and interface inversion asymmetry (IIA)  in TI quantum wells ~\cite{Tarasenko}.  
 In this calculation we restrict ourselves to standard BHZ model. The impact of the SOI on the QD electron spectra will be discussed at the end of the next section.

 We assume that the QD potential is rectangular and circularly symmetrical, $V (r) = V \theta (R-r)$, where $V$ and $R$ are depth and radius of the dot, and $\theta(x)$ is the Heaviside step function.

The total Hamiltonian of the system, $H_0 + V (r)$, contains two separate spin blocks for which the Schr\"odinger equations are:
\begin{equation}
 \begin{array}{rl}
[E\sigma_0-h(\mathbf{k})]\Psi_\uparrow\!=\!\sigma_0 V(\mathbf{r})\Psi_\uparrow\\
\![E\sigma_0-h^*(-\mathbf{k})]\Psi_\downarrow\!=\!\sigma_0 V(\mathbf{r})\Psi_\downarrow\\
 \end{array} ,
 \label{Eq_3a}
\end{equation}
where $\sigma_0$ is the $2\times2$ unit matrix, $\Psi_{\uparrow\downarrow}(\mathbf{r})$ are two-component spinors $(\psi_{1\uparrow\downarrow}(\mathbf{r}),\psi_{2\uparrow\downarrow}(\mathbf{r}))^T$. 
The wave functions should be quadratically normalized and, in particular, must vanish at $r\to\infty$.

Below we will describe only the calculations for spin-up electrons and hereafter omit the spin index $(\uparrow)$. The calculations of the spectrum and the wave functions of spin-down electrons are similar.

Let us introduce the following dimensionless variables and the parameters: $\varepsilon\!=\!E/|M|$, $k'= k \sqrt{(B/|M|)}$, $\{x',y',r'\}\!=\!\{x,y,r\}\sqrt{|M|/B}$,  $a\!=\!\mathcal{A}/\sqrt{MB}$, $d = D/B$, $r_0 = R\sqrt{|M|/B}$ and $v = V/|M|$, and omit the prime in the variables $x', y', r' , k'$ for convenience.

From  Eqs~(\ref{Eq_3a}) we find the bulk energy spectrum:
\begin{equation}
\varepsilon(k)=d k^2\pm\sqrt{\mu^2+(a^2+2\mu)k^2+k^4}
\label{Eq_4} ,
\end{equation} 
where $\mu = -1$ for TI and $\mu = + 1$ for trivial insulator.
Now we place the origin of the coordinate system at the center of the quantum dot and use the polar coordinate system, where
\begin{equation}
 \begin{array}{l}
\mathbf{k}^2 =-(\frac{d^2}{dr^2} +\frac{1}{r}\frac{d}{dr}+\frac{1}{r^2}\frac{d^2}{d\varphi^2})\\
k^\pm\!=\!k_x \pm k_y\! =\!e^{\pm i\varphi}(-i\frac{d}{dr} \pm\frac{1}{r}\frac{d}{d\varphi})\\
 \end{array} ,
 \label{Eq_4a}
\end{equation}
Using the trial wave functions in the form of $(\psi_1,\psi_2)^T e^{-\lambda r}$ one finds from  Eqs~(\ref{Eq_3a}) the roots of the secular equation
\begin{equation}
\begin{array}{l}
\!\lambda_{1,2}(\varepsilon)=\frac{\sqrt{\frac{a^2}{2}+\mu+\varepsilon d\pm\sqrt{(\frac{a^2}{2}+\mu+\varepsilon d)^2-(1-d^2)(1-\varepsilon^2)}}}{\sqrt{1-d^2}},\\
\lambda_{3,4}(\varepsilon)=\lambda_{1,2}(\varepsilon-v).\\
\end{array}
\end{equation} 

In what follows, we restrict ourselves to the case, when the parameter $a\geq 2$. 

The energy levels of the states localized in QD are located in the bulk gap of TI, ie, in the energy interval $-1 <\varepsilon <1$. If the bound energy is less than the dot bottom  ($\varepsilon<1-v$), the wavevectors $\lambda_{1,2}$ and $\lambda_{3,4}$ are real. The calculations show that in this case there are no localized states. If $1>\varepsilon>\max (1-v, -1)$ the wavevector $\lambda_4$ is imaginary, while the vectors $\lambda_{1}$, $\lambda_{2}$ and $\lambda_{3}$ are real. The wave functions of localized states must be normalized, that is they must be finite in the QD and vanish at infinity. Therefore, within the QD ($r <r_0$) they are of the form

\begin{equation}
\!\Psi_m^<(r,\varphi)\!=\!\binom{\!e^{i(m+1)\varphi}[c_3J_{m+1}(\!\lambda_3 r)\!+\!c_4 I_{m+1}(\!-i\lambda_4 r)]}{e^{im\varphi}[c_3 f_3 J_{m}(\!\lambda_3 r)\!+\!c_4 f_4 I_{m}(-i\lambda_4 r)]},\
\label{WF<}
\end{equation}
where $m$ is angular momentum quantum number. 

Outside the QD ($r> r_0$) the wave functions are as follows

\begin{equation}
\!\Psi_m^>(r,\varphi)\!=\!\binom{\!e^{i(m+1)\varphi}[c_1 K_{m+1}(\lambda_1 r)\!+\!c_2 K_{m+1}(-i\lambda_2 r)]}{\!e^{im\varphi}[c_1 f_1 K_{m}(\lambda_1 r)\!+\!c_2 f_2 K_{m}(\!-i\lambda_2 r)]}\,,
\label{WF>}
\end{equation} 
where functions $f_{i}$, ($i = 1,2,3,4$), are found by the substitution of (7) and (8) into ~(\ref{Eq_3a}) :

\begin{equation}
\!f_{1,2}\!=\!\frac{i a \lambda_{1,2}}{\mu-\!(1-d)\lambda_{1,2}^2+\varepsilon};\  f_{3,4}\!=\!\frac{i^{(3,4)} a \lambda_{3.4}}{\mu-(1\!-d\!)\lambda_{3,4}^2+\varepsilon-v}\\
\end{equation} 

Matching  the wave functions and their derivatives at the boundary $r = r_0$, one obtains a system of homogeneous linear equations for constants $c_i$ ($i = 1,2,3,4$). 
Together with the normalization condition, this system of equations determines the coefficients $c_i$.
Setting the determinant of this system equal to zero, one finds the energy spectrum of the bound states.

\section{Results and discussion\label{Results}}

The energy spectrum and, in particular, the number of states that appear in the QD depend on both the TI parameters ($a$ and $d$) and the QD parameters ($r_0$ and $v$). Here we consider in detail the case when the QD has only two levels. Such a situation is convenient for the study of physical phenomena in TI, and promising for use in quantum information systems.

\begin{figure}
\centerline{\includegraphics[width=7cm]{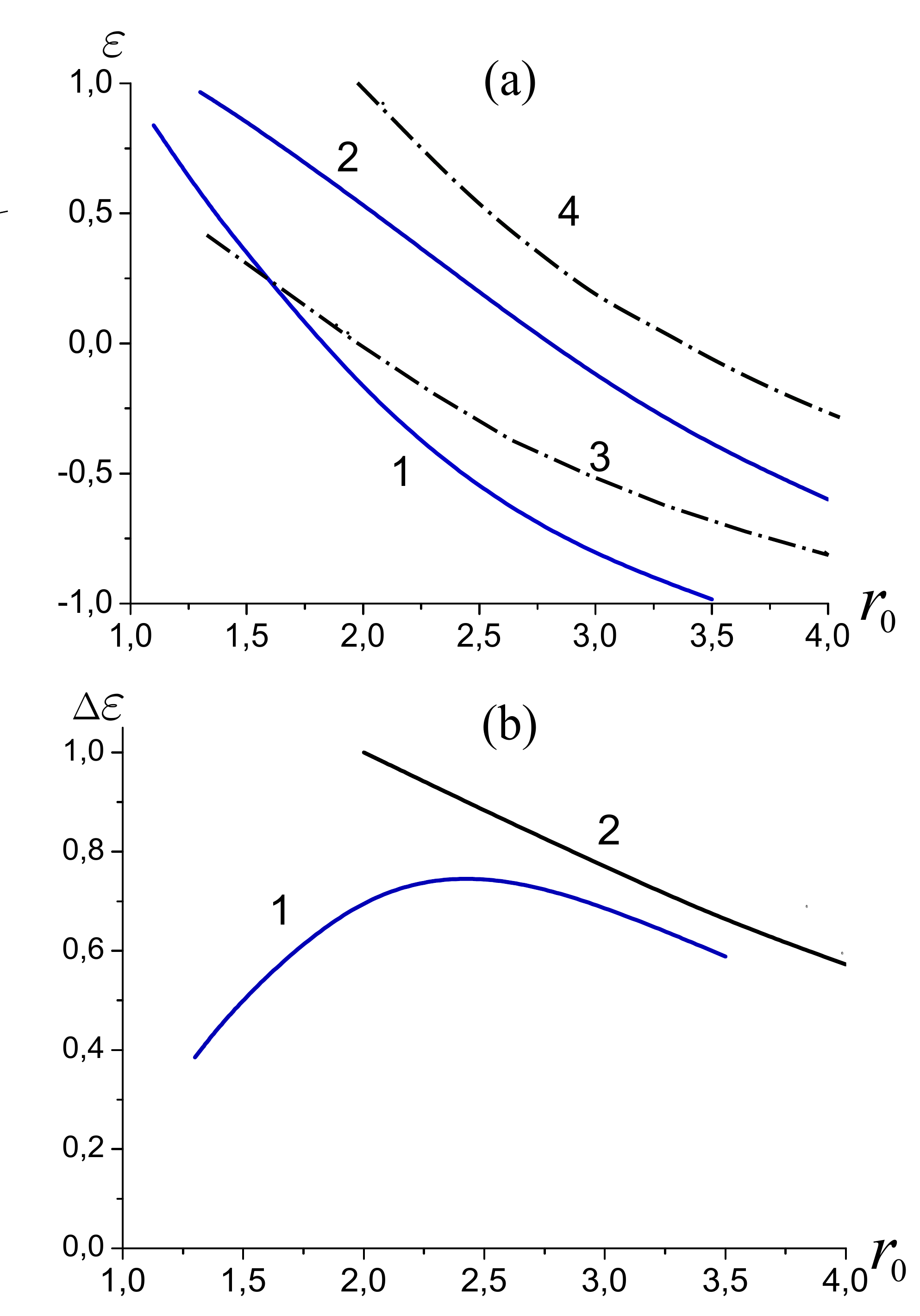}}
\caption{(Color online.) (a) The dependence of the energies of the ground and first excited states, $\epsilon_0$ and $\epsilon_1$, on the dot radius $r_0$. Blue solids lines 1 and 2 are  $\epsilon_0 (r_0)$ and $\epsilon_1(r_0)$ in the TI, black dash-dot lines 3 and 4 are similar dependences in the case of trivial insulator; (b) The dependences of the energy separation between levels of first excited and ground states on $r_0$, $\Delta \epsilon(r_0) = \epsilon_1 (r_0) - \epsilon_0(r_0)$: lines 1 (blue) and 2 (black) - $\Delta \epsilon(r_0)$  for the TI and for the trivial insulator correspondingly. The depth of the dot $|v|=2.5$, the TI  parameters are $a =2, d = 0$.}
\label{fig_E-R}
\end{figure}

Consider a QD with potential well for conduction band electrons, $v<0$. In this case the ground state is characterized by angular quantum number $m = 0$, while the first excited state has $m = - 1$. With increasing the depth $v$ (reducing the dot radius $r_0$) the energy levels of these states $\epsilon_0$ and $\epsilon_1$  fall (rise) at different speeds. Therefore, the energy separation between them changes. At high $r_0$ and small $|v|$  the separation increases with decreasing  $r_0$. However, at small  $r_0$ the situation changes and the  energy levels  $\epsilon_0$ and $\epsilon_1$  approach each other. 
This can be seen from Fig.~\ref{fig_E-R}(a), which shows the dependences of the ground and first excited states levels on the dot radius $r_0$.

From Fig.~\ref{fig_E-R} (b) one can see that the dependence of energy gap between  the excited and ground states  on $r_0$, $\Delta \epsilon(r_0) = \epsilon_1 (r_0) - \epsilon_0(r_0)$, is nonmonotonic and at small radius ($r_0 <3$) $\Delta\varepsilon$ decreases with decreasing $r_0$. 
It turns out that in deep QDs with $|v|>8$ these levels intersect when $r_0$ decreases down to $r_c < 0.7$, and then their relative position reverses.

The above situation is quite different from the case of the QD in the trivial insulator. The lines 3 and 4 in Fig.~\ref{fig_E-R}(a) and   line 2 in Fig.~\ref{fig_E-R}(b) show that in trivial phase the energies of bound states  and the energy difference between them increases with decreasing the dot radius $r_0$. This holds  for any $|v|$. As a result, at sufficiently small radius ($r_0 <2$) two bound states in QD in trivial insulator cannot coexist together for any value of the dot depth $|v|$.

The crossing of low-energy states means a change in the properties of the ground state. Indeed, when $r_0 > r_c$ the ground state  is characterized by angular momentum quantum number $m=0$, while at $r_0 < r_c$ its $m$-value is equal to $-1$.

\begin{figure}

\centerline{\includegraphics[width=8cm]{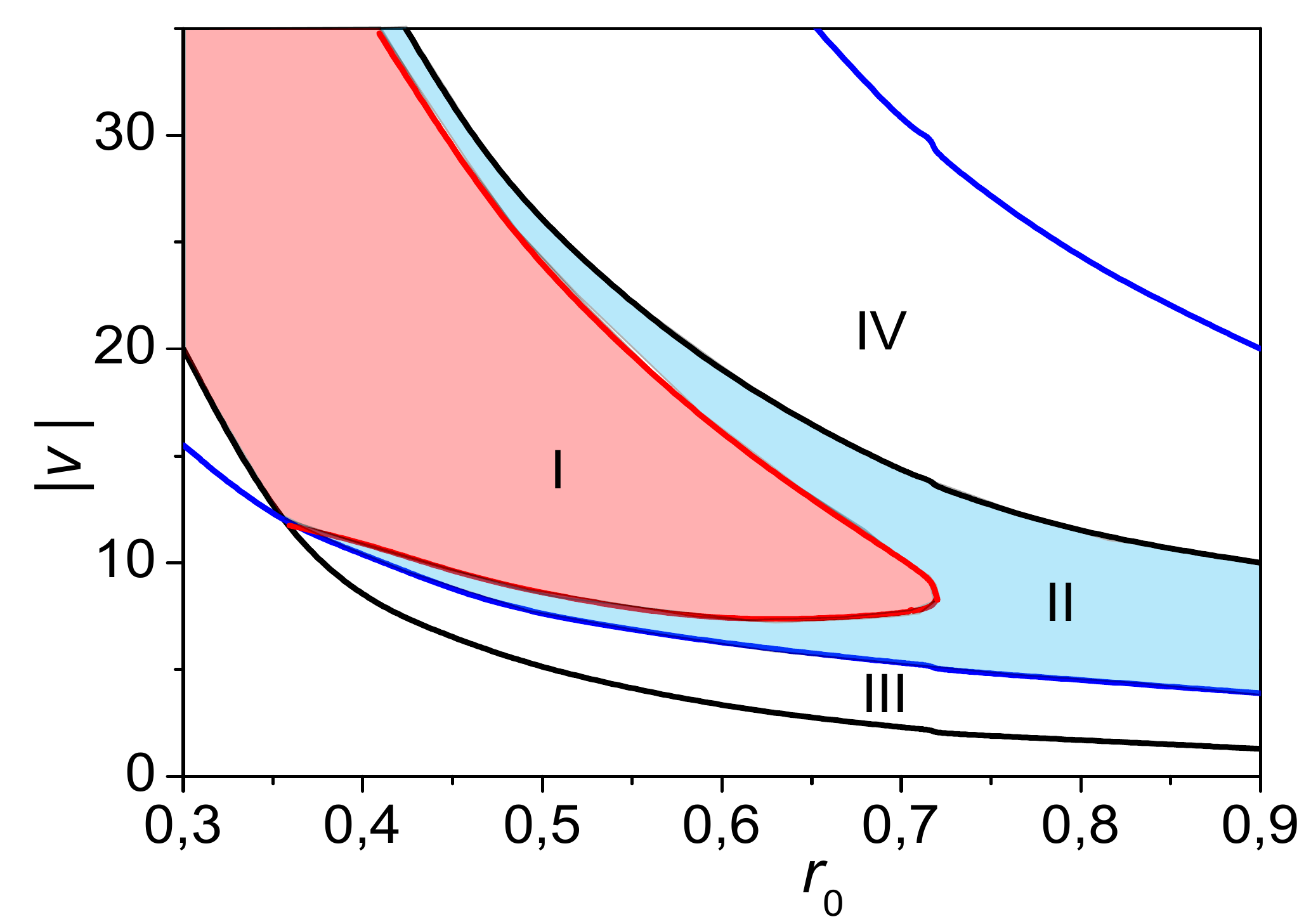}}
\caption{(Color online.) Map of bound state configurations in $v - r_0$ plane. I (red) and II (blue) are the regions where ground and first excited states levels coexist in 2D TI gap. In region II ground and excited states have angular momentum quantum numbers $m=0$ and $m=-1$, correspondingly. I is "inversed" region where ground state has $m=-1$ and excited state has  $m=0$. In regions III and IV only one of bound states can exist in the TI gap. The TI  parameters are $a =2, d = 0$.}
\label{fig_map_V-R}
\end{figure}

It is essential that level crossing and ground state reconstruction can also be induced by changing the quantum dot potential $v$. 
Fig.~\ref{fig_map_V-R} shows a map of low-energy bound states characterizing their relative positions in the 2D TI gap for different regions in a plane $(|v|, r_0)$. 
As seen from the map the increase in $|v|$ above the critical value $v_c$  for $r_0 < r_c = 0.72$ leads to a spectrum inversion, when the level of the state with $m = -1$ falls below that with $m = 0$. Interestingly,  further increase in $|v|$ results in reverse rearrangement.

The map in Fig.~\ref{fig_map_V-R}  shows that the crossing of the energy levels occurs at $r_0 = r_c = R_c \sqrt {| M |/B}\approx 0.7$.
Hence, setting  $a=\mathcal{A}/\sqrt {B | M |}=2$ , we obtain the critical size of the quantum dot $2 R_c = 4 B/\mathcal{A} \approx 20\ nm$ at $B = 0.3$ $eV nm^2$ (which corresponds to the effective mass of carriers $m * = 0.1\ m_0$,  $m_0$ - free electron mass)  and $\mathcal{A} = 4\cdot 10^{-2} \ eV nm$ ~\cite{Hu}.
 
It should be noted that at small $r_0$ the crossing of higher energy state levels with $|m| > 1$ is also possible. This occurs when the values of $v$ are several times greater than the critical value $v_c$ . 

Crossing of the low-energy levels  in the narrow QDs manifests itself in the dependencies of these levels on the parameters $M$ and $\mathcal{A}$ as well.
Interestingly, in bilayer quantum structures such as InAs/GaSb where inverted band spectrum is created by the hybridization of electronic states of a one quantum well (InAs) and hole states of other quantum well (GaSb) ~\cite{Hu,Suzuki,Qu}, the TI parameters $M$ and $\mathcal{A}$ can be easily changed by transverse voltage. This provides an efficient way to control the energy level crossing.

Bound states also arise in the quantum dots produced by a positive potential, $v> 0$. When parameter $d$ which determines the asymmetry of the electron and hole branches of energy spectrum is equal to zero, the energy levels of the bound states, counted up from the valence band top depend on $r_0$, $v$ and other parameters in the same manner as described above. However, in the case where the parameter $d$ is not small ($0.1<d<1$ ), the spectra generated by potentials of opposite signs are essentially different.
With increasing $d$ the conditions for the inversion of two low-lying states are relieved for $v> 0$ but become tougher for $v <0$. In the first case, the value of $r_c$ increases up to 0.9, while $v_c$ drops down to 3 with an increase of $d$ up to 0.8.

The crossing of  $\epsilon_0$ and $\epsilon_1$ levels leads to interesting features in the  optical transitions between the quantum dot levels.

 In accordance with the selection rules for electric dipole transitions the absorption of the right-hand (left-hand) circular polarized  light leads to an increase (decrease) in orbital quantum number $m$ by 1 (-1).
 Therefore, in the normal arrangement of the levels, when $\epsilon_0(m = 0) <\epsilon_1 (m = -1)$,  left-hand circularly polarized light  is absorbed, while after change of level positions (when $\epsilon_0(m = 0) >\epsilon_1 (m = -1)$) the optical transitions in TI QD  are  induced by right-hand circularly polarized light.

The matrix element of the optical dipole transition 
\begin{equation}
\mathcal{M}=\!\int_0^{\infty}\!rdr\int_0^{2 \pi}\!d\varphi\langle\Psi_f^*(r,\varphi)|h(\mathbf{A})|\Psi_i(r,\varphi)\rangle,
\label{matrix_elem}\\
\end{equation}
is determined by the operator of the interaction between the light and 2D TI electrons
\begin{equation}
h(\mathbf{A})=
\begin{pmatrix}
 \!2(B\!+\!D)\mathbf{k}\mathbf{A} & \mathcal{A}(A_x+iA_y)\\
 \mathcal{A}(A_x-iA_y) & 2(B\!-\!D)\mathbf{k}\mathbf{A} 
\end{pmatrix}\,,
\label{H_A}
\end{equation} 

Using the found energy of the bound states, the wave functions $\Psi_i (r, \varphi)\!=\!\Psi_{m = 0}(r,\varphi)$ and $\!\Psi_f(r,\varphi)\!= \!\Psi_{m = -1}(r,\varphi)$, and the expressions for operators $k_x = (k^+ +k^-) /2$ and $k_y = (k^+- k^-)/2i$ ~\ref{Eq_4a} we calculate the matrix elements.

\begin{figure}
\centerline{\includegraphics[width=8cm]{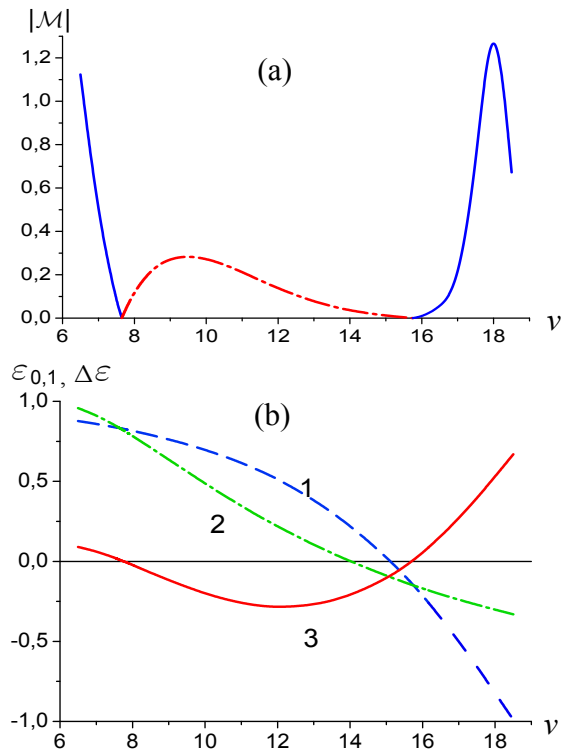}}
\caption{(Color online.) (a) The dependence of the modulus of the matrix element of the dipole optical transition between states $\epsilon_0$ and $\epsilon_1$ on the depth of the potential well, $|\mathcal{M}|(v)$. The solid blue lines (red dash-dot line) correspond to absorption of the left-hand (right-hand) circularly polarized light (b)  The dependence of the energies of low-laying states $\epsilon_0$ (blue dash line), $\epsilon_1$ (green dash-dot) and energy difference $\Delta \epsilon = \epsilon_1- \epsilon_0$ (red solid) on the potential $v$. The TI parameters: $a=2$ and $d = 0$.}
\label{fig_I(V)}
\end{figure}

Fig.~\ref{fig_I(V)}(a) and Fig.~\ref{fig_I(V)} (b) show the dependences of the modulus of the optical transition matrix element    $\mathcal{M}$,  the dependences of the energy levels $\epsilon_0$ and $\epsilon_1$ (ground and excited states for $v< 8$) and the difference in their energy $\Delta \epsilon = \epsilon_1- \epsilon_0$ on the dot depth $v$.

One can see that with the increase of $v$ the direction of circular polarization of absorbed light changes from the left-hand circular polarization to the right-hand one,
and then back again when the levels $\epsilon_0$ and $\epsilon_1$  cross once more.
Of course, the interaction of the light with electrons with opposite spin ($\downarrow$)  leads to the similar features in the absorption spectrum, but with opposite direction of circularly polarized light, and this results in full compensation of polarization effects.

However, the spin degeneracy, ie, the degeneracy of the states of the electrons of different spin blocks is removed when the time reversal symmetry is broken due to the presence of  an external magnetic field or the exchange interaction with magnetized localized spins. The latter case is realized in magnetic topological insulators (MTI) with out-of-plane magnetization ~\cite{Liu_Zhang,Yu,Wang}. In this case,  the exchange term $H_{ex}$, that induces the Zeeman splitting of TI bands, should be added to the Hamiltonian (1) 
\begin{equation}
 H_{ex}=
\begin{pmatrix}
 G_e & 0\\
 0 &G_h
\end{pmatrix}\,,
\label{HH}
\end{equation} 
where $2G_e$ and $2G_h$ are the spin splitting for conduction and valence bands, respectively.
The total Hamiltonian of MTI, $H^* = H + H_{ex}$, has a block-diagonal form as before.

 The energy spectrum of each block is shifted by the value $\Delta = \pm (1/2) (G_e + G_h)$, and the bulk gap changes by $\Delta M =\pm (G_e - G_h)$, where the plus sign (+) refers to spin-up electrons and the sign (-) to spin-down electrons. 

Procedure  of the bound states spectra calculations does not change. The calculation results can be obtained by proper replacement of gap parameter $M$ by different parameters $M_{\uparrow, \downarrow}$ for each of two spin blocks. 
 In the simple case, when the $G_e = - G_h = -G$, the shift of the energy gap center is absent ($\Delta = 0$), and the energy gap for spin-up electrons  increases  in absolute value with increasing $G$, $M_\uparrow = - |M| - 2 G$, while for electrons of spin-down block  it decreases, $M_\downarrow = - |M| + 2 G$.

 When $2 G> |M|$ , the MTI is in quantum anomalous spin Hall  phase,  where the electrons with spin-down are in the normal phase. Then only the states of the electrons with spin-up  can participate in the optical transitions in deep narrow QDs. So in this case the level crossing results in the change of the direction of    circular polarization of absorbed light.

Now let's turn to the possibility of observing  the level crossing experimentally.
Apparently, the spectrum reconstruction in 2DTI quantum dots can be detected and  studied by the changes in the intensity and polarization of the absorption of light passing  through 2D TI with quantum dots.
In this respect, it is highly interesting to study  the transient electroluminescence, which must occur at the intersection of the QD levels  under the pulse voltage application. Voltage induced inversion of level positions leads to the violation of their equilibrium population (and even to the inversion of electron function distribution) and therefore to the spontaneous emission of polarized light, which results in the recovery of equilibrium electron distribution in  QD.
 When the voltage is turned off, there is a return to the previous configuration level positions, and again the electron population of the exited state is higher than the equilibrium one, leading to a new peak light emission (but with a different circular polarization).

Importantly,  because of the Coulomb repulsion  between electrons, only one of the states (the ground or excited state)  can be filled by electron, if the energy levels of these states lie below the Fermi level not more than 
$\Delta E= e^2/{\epsilon^0\lambda}$, where $\lambda$ is the size of electron localization and $\epsilon^0$ is relative permittivity. For $\lambda=10 nm$ and $\epsilon^0 = 10 \div 100$ one finds  $\Delta E= (1 \div 10) meV$, this value is comparable to the typical energy gap of 2D TI.

Note that the level crossing can also appear in the  spectrum of bound states of point defects in the 2D TI bulk ~\cite{Sabl,Sablik}. The above described features of the optical properties of  QD are inherent to this case also. The reconstruction of the point defect spectrum can be induced by electrical tuning the 2D TI parameters $M$ and $\mathcal{A}$.

 It should be also emphasized  that in real TI structures, there are  two factors that can affect the QD energy spectra, namely the SOI and the hybridization of QD bound states with the edge states. A detailed study  of these factors  requires a separate research. However, to understand the impact of these factors on the QD spectra one can use the bound state calculations in model of defect with a short-range potential~\cite{Sabl, Sablik,Sablik2}. As noted previously, such point-like defects can be regarded as  model quantum dots of small radius. 

The SOI due to  BIA~\cite{Koenig}  and IIA  in TI quantum wells ~\cite{Tarasenko} is described by adding the term $\hat\tau_y \otimes \hat\sigma_y \Delta_s$,  where $\Delta_s$ is SOI parameter, to BHZ Hamiltonian.  
The calculations show that the SOI has only a minor effect on the QD spectra for $\Delta / M < 0.2 $ ~\cite{Sablik}. However, the SOI is essential near the level crossing leading to avoided crossing (anticrossing). For   $\Delta_s / M = 0.2$ the anticrossing gap is about $ 0.05 M$.    

In the case of magnetic TIs  the anticrossing is strongly suppressed  due to  the spin degeneracy lifting .  When the magnetization is increased up to $ |G| = 2|M|$,  the  anticrossing gap decreases by almost one order of magnitude.

The hybridization of the bound states with the edge states for case of point-like defect was studied in ~\onlinecite{Sablik2}. It leads to the transformation of the localized states to the resonances. The spectra of edge states do not change, but the local electronic density of states at the resonance energy has a sharp peak near the defect. The hybridization  decreases with increasing distance between the defect and TI edge $l$ and vanishes at $l \gg \sqrt{M/B} $.

\section{Conclusion\label{Summary}}
In conclusion, we first briefly  discuss the reasons for levels crossing in the energy spectrum of 2D TI quantum dot and then summarize the main results.

The energy of the deep bound states in QD is determined not only by the size quantization but by the hybridization of electron and hole bands as well. In shallow QDs (at $|V|<<2| M |$) the bound states wave functions are formed by wave functions of  conduction band. The ground state has minimal orbital energy and is characterized by angular momentum quantum number $m = 0$ and its electron density $n (r)$ is concentrated in the dot center. The first excited state has $m = -1$, with the density $n (r)$, being concentrated at the QD edge. For small $V$ ($|V|\!<<\!2|M |$) the increase of $V$ and the decrease in $R$  lead to the increase in the energy separation between the levels of these states due to the growth of size quantization energy. 

Further increase in $|V|$ induces an admixture of valence band states to the  bound states. According to  Eqs.(~\ref{WF<}) and (\ref{WF>}),  the  admixture creates the electron spinor components  having the orbital quantum numbers $m = 1$ and $m = 0$ in ground and excited states respectively. The electron density with $m = 0$ has smaller orbital quantization energy than the one with $m \ne 0$.  That is why with increasing $|V|$ the excited state energy decreases more rapidly than the energy of the ground state. As a result the levels approach each other and intersect at sufficiently high  $|V|$  and small $R$. 

In trivial insulator this trend is fully canceled by an increase in the radial quantization energy,  which in its turn increases with $|V|$. This is due to the fact that the radial  quantization energy in trivial insulator is larger than that in TI, because it is determined by lower electron effective mass. Indeed, determining the electron effective mass $m$ as the inverse of the coefficient by $k ^ 2$ in the electronic spectrum $\varepsilon(k)$  we find from Eq.~(\ref{Eq_4}) $m_{T} \sim 1 /( a ^ 2 - 2) $ for TI and  $m_{t}\sim 1/( a ^ 2 + 2) $  for the trivial insulator. For a = 2 we have $m_{T}/m_{t}  = 3$. 
Finally, we summarize the main results.

Non-trivial  features of 2D TIs can manifest itself not only in the emergence of robust edge states at the boundaries, but also in the unusual spectra of bound states at defects and  in microstructures.

 In this regard, we have studied the spectra of QDs formed in 2D TI by the electrostatic potential. 
It turned out that some energy levels vary with the depth of the potential well in an unusual way. In QDs of small size (less than the critical one) the energy gaps between  some levels, change non-monotonically, that can lead to their intersections. The crossing of the lowest energy states is accompanied by a radical change of the groundstate wave function. 

 It is important that the level crossing can be controlled by tuning of the topological insulator parameters by external transverse voltage.

The results show the major role of interband hybridization in the formation of the energy spectrum of bound states in 2D TI quantum dots. 

The  spectrum reconstruction results in a non-trivial changes in the optical properties of QDs in 2D TI.
In particular,  the level crossing  leads to: i) the significant changes in light absorption intensity, ii) the efficient electroluminescence caused by electrical tuning 
the QD and TI parameters, and iii)  the changes in the polarization direction of the absorbed circularly-polarized light in the case of  magnetic 2D TI.

We believe that the measurements of the optical properties of  QDs in TI can allow one to detect the level crossing, to study an unusual rearrangement of the bound state spectra    and thus  to obtain the better understanding of 2D TI physics.

\section*{Acknowledgments}
Author thanks V.~A. Sablikov for many stimulating discussions.
This work was supported by Russian Science Foundation under the grant No. 16-12-10335.

\end{document}